\documentclass[%
reprint,
superscriptaddress,
amsmath,amssymb,
aps,
]{revtex4-2}

\usepackage{graphicx}% Include figure files
\usepackage{dcolumn}% Align table columns on decimal point
\usepackage{bm}% bold math
\usepackage{hyperref}% add hypertext capabilities
\usepackage{pgf}
\usepackage{soul}  % to highlight some changes

\usepackage{times} %change font to times new roman
\usepackage{siunitx} % for SI units
\usepackage{physics} %%% braket notation  etc.
\begin{document}

\title{Redirection and Reshaping of Intense Extreme-Ultraviolet Radiation}

\author{Yu He}
\email{yuhe@mpi-hd.mpg.de}
\affiliation{Max-Planck-Institut für Kernphysik, Saupfercheckweg 1, 69117 Heidelberg, Germany}

\author{Alexander Magunia}
\affiliation{Max-Planck-Institut für Kernphysik, Saupfercheckweg 1, 69117 Heidelberg, Germany}

\author{Harijyoti Mandal}
\affiliation{Max-Planck-Institut für Kernphysik, Saupfercheckweg 1, 69117 Heidelberg, Germany}

\author{Muwaffaq Ali Mourtada}
\affiliation{Max-Planck-Institut für Kernphysik, Saupfercheckweg 1, 69117 Heidelberg, Germany}

\author{Carlo Kleine}
\affiliation{Max-Planck-Institut für Kernphysik, Saupfercheckweg 1, 69117 Heidelberg, Germany}

\author{Arikta Saha}
\affiliation{Max-Planck-Institut für Kernphysik, Saupfercheckweg 1, 69117 Heidelberg, Germany}

\author{Marc Rebholz}
\affiliation{Max-Planck-Institut für Kernphysik, Saupfercheckweg 1, 69117 Heidelberg, Germany}

\author{Gergana D. Borisova}
\affiliation{Max-Planck-Institut für Kernphysik, Saupfercheckweg 1, 69117 Heidelberg, Germany}

\author{Lina Hedewig}
\affiliation{Max-Planck-Institut für Kernphysik, Saupfercheckweg 1, 69117 Heidelberg, Germany}

\author{Hannes Lindenblatt}
\affiliation{Max-Planck-Institut für Kernphysik, Saupfercheckweg 1, 69117 Heidelberg, Germany}

\author{Florian Trost}
\affiliation{Max-Planck-Institut für Kernphysik, Saupfercheckweg 1, 69117 Heidelberg, Germany}

\author{Ulrike Frühling}
\affiliation{Deutsches Elektronen-Synchrotron DESY, Notkestraße 85, 22607 Hamburg, Germany}

\author{Christina C. Papadopoulou}
\affiliation{Deutsches Elektronen-Synchrotron DESY, Notkestraße 85, 22607 Hamburg, Germany}

\author{Elisa Appi}
\affiliation{Department of Physics, Lund University, SE-221 00 Lund, Sweden}

\author{Stefan Düsterer}
\affiliation{Deutsches Elektronen-Synchrotron DESY, Notkestraße 85, 22607 Hamburg, Germany}

\author{Tino Lang}
\affiliation{Deutsches Elektronen-Synchrotron DESY, Notkestraße 85, 22607 Hamburg, Germany}

\author{Skirmantas Alisaukas}
\affiliation{Deutsches Elektronen-Synchrotron DESY, Notkestraße 85, 22607 Hamburg, Germany}

\author{Christoph M. Heyl}
\affiliation{Deutsches Elektronen-Synchrotron DESY, Notkestraße 85, 22607 Hamburg, Germany}
\affiliation{GSI Helmholtzzentrum für Schwerionenforschung GmbH, Planckstraße 1, Darmstadt 64291, Germany}
\affiliation{Helmholtz Institute Jena, Fröbelstieg 3, Jena 07743, Germany}

\author{Steffen Palutke}
\affiliation{Deutsches Elektronen-Synchrotron DESY, Notkestraße 85, 22607 Hamburg, Germany}

\author{Markus Braune}
\affiliation{Deutsches Elektronen-Synchrotron DESY, Notkestraße 85, 22607 Hamburg, Germany}

\author{Christina Bömer}
\affiliation{Deutsches Elektronen-Synchrotron DESY, Notkestraße 85, 22607 Hamburg, Germany}

\author{Dietrich Krebs}
\affiliation{Deutsches Elektronen-Synchrotron DESY, Notkestraße 85, 22607 Hamburg, Germany}

\author{Doriana Vinci}
\affiliation{European XFEL, Holzkoppel 4, 22869 Schenefeld, Germany}

\author{Philip Mosel}
\affiliation{Institut für Quantenoptik, Leibniz Universität Hannover, Welfengarten 1, 30167 Hanover, Germany}

\author{Peer Biesterfeld}
\affiliation{Institut für Quantenoptik, Leibniz Universität Hannover, Welfengarten 1, 30167 Hanover, Germany}

\author{Ingmar Hartl}
\affiliation{Deutsches Elektronen-Synchrotron DESY, Notkestraße 85, 22607 Hamburg, Germany}

\author{Robert Moshammer}
\affiliation{Max-Planck-Institut für Kernphysik, Saupfercheckweg 1, 69117 Heidelberg, Germany}

\author{Milutin Kovacev}
\affiliation{Institut für Quantenoptik, Leibniz Universität Hannover, Welfengarten 1, 30167 Hanover, Germany}

\author{Kiyoshi Ueda}
\affiliation{Max-Planck-Institut für Kernphysik, Saupfercheckweg 1, 69117 Heidelberg, Germany}
\affiliation{Department of Chemistry, Tohoku University, Sendai 980-8578, Japan}
\affiliation{Center for Transformative Science and School of Physical Science and Technology, ShanghaiTech University, Shanghai 201210, China}

\author{Mette B. Gaarde}
\affiliation{Department of Physics and Astronomy, Louisiana State University, Baton Rouge, Louisiana 70803, USA}
\author{Christian Ott}
\email{christian.ott@mpi-hd.mpg.de}
\affiliation{Max-Planck-Institut für Kernphysik, Saupfercheckweg 1, 69117 Heidelberg, Germany}

\author{Thomas Pfeifer}
\email{thomas.pfeifer@mpi-hd.mpg.de}
\affiliation{Max-Planck-Institut für Kernphysik, Saupfercheckweg 1, 69117 Heidelberg, Germany}

%\date{\today}

\begin{abstract}

The goal to control short-wavelength radiation for the investigation and manipulation of ultrafast dynamics in quantum systems coevolves with the growing availability of extreme-ultraviolet (XUV) and x-ray sources from high-harmonic generation and free-electron lasers. Here, we present an XUV spatio-spectral phase modulator based on an intense XUV laser beam propagating through an optically thick resonant target, introducing dispersion profile variations around the resonance both perpendicular to and along the laser propagation direction. The resulting dipole radiation gets spectrally reshaped and becomes more divergent as compared to the original beam in the far field. As an experimental demonstration, the intense-XUV-induced double-peak off-axis structure in the far-field spectrum obtained at the Free-Electron Laser in Hamburg (FLASH) shows indications of the underlying XUV-driven Rabi dynamics and resonant pulse propagation effects. The presented work highlights a ubiquitous phenomenon occurring when an intense laser beam passes through a resonant medium.

\end{abstract}

\maketitle 
Ultrashort laser pulses with high intensity have widespread applications in modern society from fundamental research to advanced medical and industrial applications. The interaction of matter with strong laser fields has revealed a wealth of nonlinear phenomena such as ac Stark shift, self-phase modulation, multiphoton ionization, and high-harmonic generation \cite{RevModPhys.72.545}. The extension of nonlinear spectroscopy from optical to the extreme-ultraviolet (XUV) and x-ray region enables novel insights into electronic dynamics with unprecedented temporal and spatial resolution \cite{Young_2018, chergui2023progress}. Recent technical developments in XUV and x-ray free-electron lasers (FELs) at large-scale facilities providing ultrashort and high-brilliance light pulses with tunable wavelengths have opened new research opportunities previously out of reach \cite{young2010femtosecond, rudek2012ultra, PhysRevLett.110.173005, PhysRevLett.111.233902, RevModPhys.88.015006, PhysRevLett.120.263901, PhysRevLett.123.163201, PhysRevLett.123.103001,PhysRevX.11.031001,nandi2022observation, doi:10.1126/sciadv.adk1482, doi:10.1126/science.adn6059, richter2024strong}.

\begin{figure}[ht]
\includegraphics[width=0.5\textwidth]{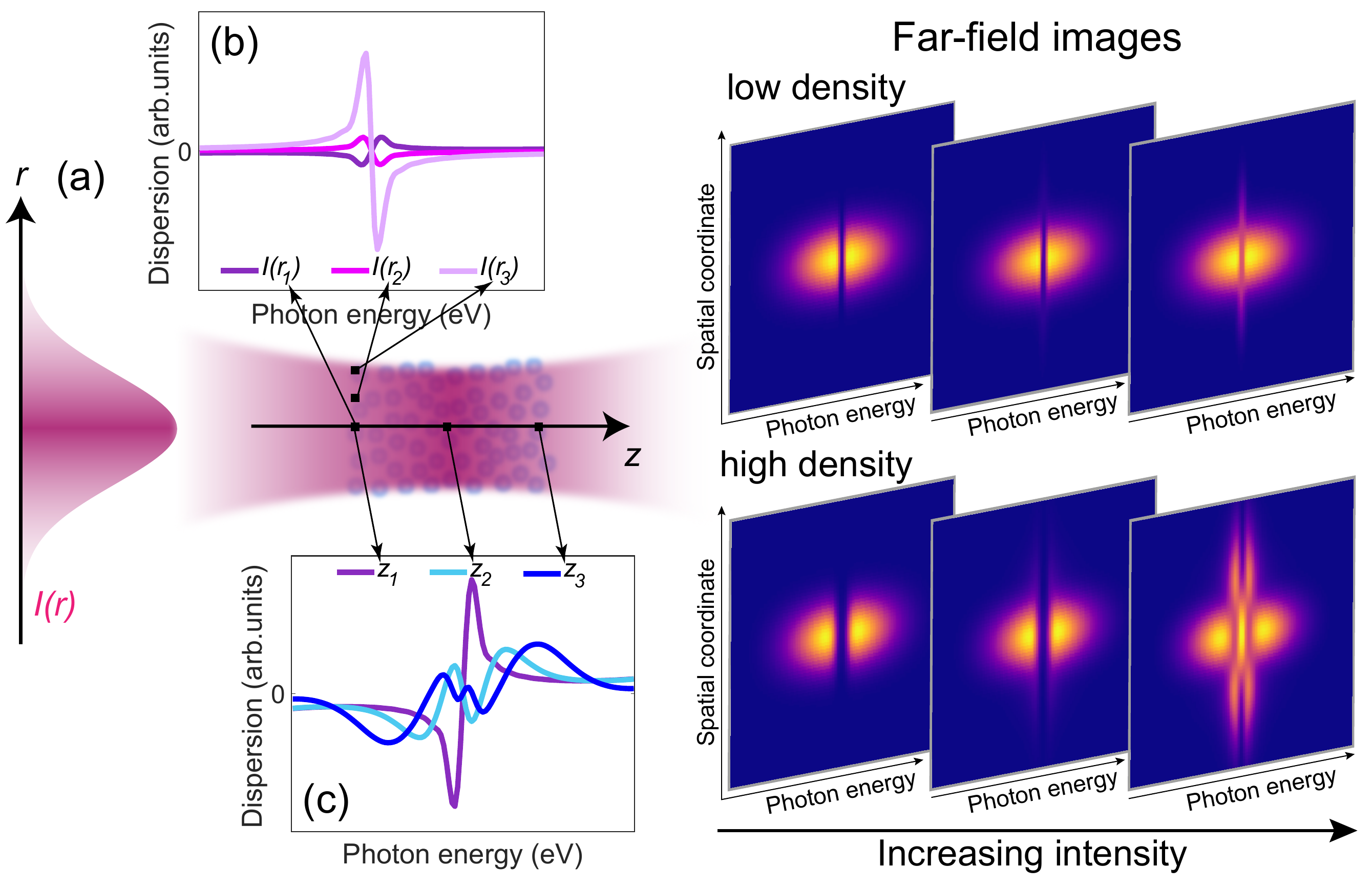}
\caption{\label{fig1} \textbf{Concept of the XUV spatio-spectral phase modulator.} (a) A focused XUV beam (cylindrical symmetry is assumed) passes through a target medium and imprints a spectral hole at the transition energy in the transmitted spatio-spectral profile (spatial coordinate: $x$ or $y$ in Cartesian coordinate). When the pulse intensity is high enough, the refractive index (dispersion) of the target in the vicinity of the resonance gets modified due to the nonlinear interaction. This modification depends on the pulse intensity and hence varies along the transverse direction [panel (b)]. The medium thus imparts a nonuniform phase shift to the XUV beam, thereby causing wavefront curvature and redirection of the resonant spectrum. For a medium with a higher target density, macroscopic pulse propagation manifests itself in further modification of the dispersion profiles [panel (c)] and the reshaping of the resonant spectral structures. The dispersion profiles in panels (b) and (c) are shown in arbitrary units for better visibility.}
\end{figure}

\begin{figure*}[ht]
\includegraphics[width=0.95\textwidth]{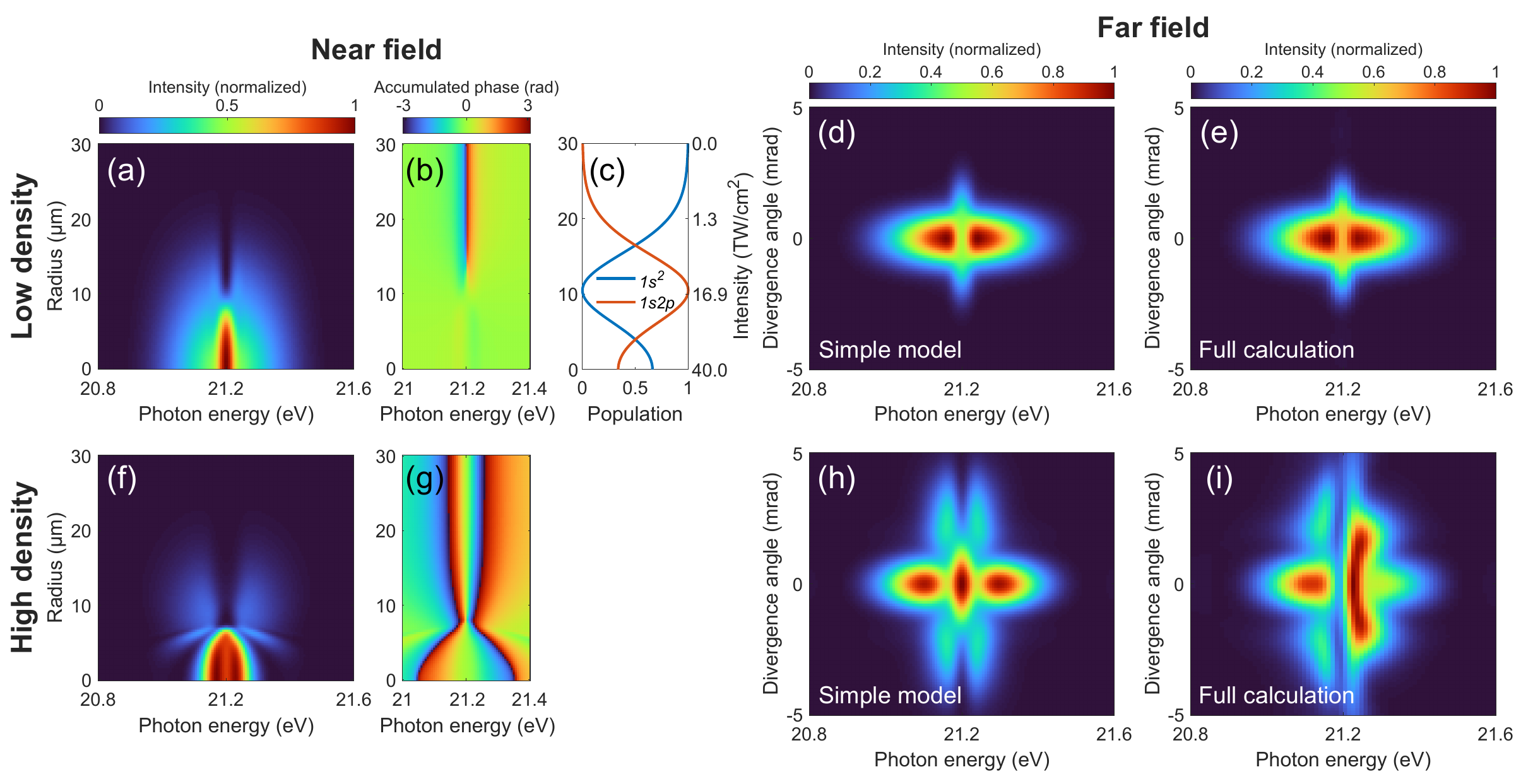}
\caption{\label{fig2} \textbf{Model simulation of the intense-XUV-induced spatial redirection and spectral reshaping in optically thin and thick media.} (a, f) Near-field radial-spectral profile, (b, g) accumulated phase of the XUV beam at the exit of the medium, and the resulting (d, h) far-field spatio-spectral profiles. (c) Evolution of the state populations after the interaction with the incoming pulse as a function of radial distance from the beam center (i.e., peak XUV pulse intensity). (e, i) Far-field spatio-spectral profiles obtained from the full TDSE-MWE simulations. The atomic densities in panels (a-e) and (f-i) are $2.4\times10^{16}$ and $1.9\times10^{17}$ $\si{cm^{-3}}$, respectively. The evolution of the state populations in the high-density case depends also on the propagation distance $z$ as the driving XUV pulse undergoes temporal reshaping, and panel (c) essentially represents the results at the entrance.}
\end{figure*}

The absorption and dispersion of light are omnipresent phenomena in nature. They are connected to the complex refractive index of the interacting medium and can be microscopically described in terms of the induced oscillating electric dipoles. Following the excitation of the system, free induction decay produces reemitted radiation at the resonance frequency \cite{PhysRev.70.460,PhysRev.77.297.2}. By changing the phase and amplitude of the dipole emitters with an external laser field, attosecond transient absorption spectroscopy has enabled the control over the absorption line shapes \cite{PhysRevLett.105.143002, Ott716, PhysRevLett.112.103001, Kaldun738, Cao_2016, PhysRevA.95.043427, PhysRevLett.121.173005}. Going beyond the dilute-gas limit, the collective light-matter interaction resulting from the macroscopic pulse propagation inside the medium comes into play \cite{PhysRevA.1.1604, RevModPhys.43.99}, which gives rise to spectral reshaping near the resonance \cite{PhysRevA.88.051402, PhysRevLett.114.143002, PhysRevA.93.033405, PhysRevA.83.013419, PhysRevA.87.013415, PhysRevA.88.033409, PhysRevA.102.053113, PhysRevA.105.063515, PhysRevA.103.L041102, PhysRevLett.129.273201, PhysRevResearch.6.013103}. In addition to altering the absorption properties, an interesting situation arises when the dispersion profile gets modified. Since the wave vector $\bm{k}$ follows the expression
\begin{equation}
\label{wavevector}
\bm{k}(\omega,r,z)= \bm{\nabla}\varphi(\omega,r,z), 
\end{equation}
any inhomogeneity in spectral phase $\varphi$ accumulated due to dispersion along the transverse direction $\bm{r}$ perpendicular to the incident laser propagation direction $\bm{z}$ would result in the spatial redirection of the beam. This has been demonstrated in the weak XUV regime by virtue of a spatially offset auxiliary infrared beam \cite{bengtsson2017space, Bengtsson_2019} or a gas jet with a transverse density gradient \cite{drescher2018extreme}. However, it has not been demonstrated for intense short-wavelength radiation.

In this work, we demonstrate the spatial redirection and spectral reshaping of intense XUV radiation by passing solely a single XUV beam through a resonant medium. This is achieved by the self-induced change of the dispersion profile in the vicinity of an atomic resonance through the interplay of nonlinear light-matter interaction and resonant pulse propagation, thus modifying the spectral phase accumulation in Eq.~(\ref{wavevector}) into $\varphi[\omega,r,I(r),z]$. The schematic principle of the concept is presented in Fig.~\ref{fig1}. A focused XUV beam propagates through the medium, which acts as a spatio-spectral phase modulator. When the interaction is in the nonperturbative regime, a frequency-dependent modification of the refractive index around the resonance is introduced, which is intensity-dependent and hence varies along the radial direction as shown in Fig.~\ref{fig1}(b), leading to more divergent radiation close to the resonance as compared to the original beam in the far field. For a medium with a higher target density, the refractive index gets further distorted along the propagation direction [Fig.~\ref{fig1}(c)] as the driving pulse undergoes temporal reshaping, giving rise to the appearance of more spectral structures near the resonance position.

This Letter is organized as follows: We start with a simple computational model to identify the characteristic features of the spatial redirection and spectral reshaping of an intense laser beam propagating through macroscopic media. The robustness of the effect is further supported by a large-scale numerical calculation. Finally, the intense-XUV-FEL-induced results are presented as an experimental demonstration of the concept.

The considered model system consists of the ground state $1s^2$ and the singly excited state $1s2p$ of helium, and is subjected to an ultrashort XUV pulse $\mathcal{E}(t)$. The wave function of the system can be written as $\ket{\Psi(t)} = c_{1}(t)\ket{1s^2} + c_{2}(t)e^{-i\omega_{12}t}\ket{1s2p}$, with $\omega_{12}$ denoting the transition frequency. The evolution of the state coefficients $c_{1}(t)$ and $c_{2}(t)$ are computed by solving the time-dependent Schrödinger equation (TDSE). The spatial evolution of the XUV spectrum in one dimension along the propagation direction is approximated by \cite{PhysRevA.83.033405}
\begin{equation}
\label{waveequation}
\frac{\partial}{\partial z}\tilde{\mathcal{E}}(\omega,z) = -2\pi i \frac{\omega}{c} \tilde{P}(\omega,z),
\end{equation}
which was formulated in a reference frame moving with the XUV laser pulse at the speed of light in vacuum $c$. The polarization response in the frequency domain $\tilde{P}(\omega,z)$ is related to the dipole spectrum $\tilde{d}(\omega,z)$ by $\tilde{P}(\omega,z) = 2N\tilde{d}(\omega,z)$, in which $N$ denotes the atomic number density and 2 accounts for the response of two electrons in helium. Here, we decouple the XUV radial and spectral evolution to isolate the effect induced by different pulse intensities: The focusing and defocusing of the XUV beam within the medium is neglected, and each point along the radial direction is treated independently with different input peak intensities following a Gaussian distribution. In this case, any induced inhomogeneity in spectral phase $\varphi$ along the radial direction in Eq.~(\ref{wavevector}) originates from the nonuniform radial intensity distribution. To convert the transmitted XUV spectrum from the near field to the far field where it is actually recorded in the experiment, a Hankel transformation is further performed to account for the propagation of the XUV beam in free space \cite{lin2018attosecond}.

We choose a cosine-squared XUV pulse of 8 fs full width at half maximum (FWHM) duration, centered on resonance at 21.2~eV, with a peak intensity of 40~$\rm{TW/cm^{2}}$. These pulse parameters are chosen to approximate the dominant spike contained within the FEL pulse in the experiment shown below. The dipole matrix element between the two states is 0.3~a.u.. We consider a constant Gaussian beam radius of 15~$\si{\micro m}$ propagating through the target medium of 2 mm length. The calculated radial-spectral profile of the XUV beam at the exit of the medium is shown in Fig.~\ref{fig2}(a), for a rather dilute medium with an atomic number density of $2.4\times10^{16}$ $\si{cm^{-3}}$, where the pulse reshaping effect is insignificant and the results are similar to the single-atom prediction. With decreasing radial distance towards the center of the beam, corresponding to increasing peak pulse intensity, the spectral profile in the vicinity of the resonance in the transmitted spectrum changes from a dip to a peak. This spectral lineshape change is accompanied by a spectral phase change as shown in Fig.~\ref{fig2}(b), and is associated with the significant population transfer between the states as shown in Fig.~\ref{fig2}(c). The returning electrons to the ground state after completing half a Rabi cycle carry an additional phase of $\pi$, thus turning the natural absorption of the resonance into emission. The Rabi oscillations driven by intense short-wavelength light has received considerable interest in recent years \cite{PhysRevLett.123.163201, nandi2022observation, PhysRevLett.131.043201, richter2024strong} and is a signature of the nonlinear XUV light-matter interaction. The resulting spatial phase variation along the radial axis leads to a redirected XUV emission in the far field at the resonant frequency, which is plotted in Fig.~\ref{fig2}(d). For a higher atomic density of $1.9\times10^{17}$ $\si{cm^{-3}}$, the collective light-matter interaction within the medium leads to strong temporal and spectral pulse reshaping \cite{PhysRevA.88.051402, PhysRevLett.114.143002, PhysRevA.93.033405}. The peak in the near-field spectrum develops into two as presented in Fig.~\ref{fig2}(f), accompanied by a broadened spectral phase profile as shown in Fig.~\ref{fig2}(g), which translates into two off-axis emission peaks in the far field in Fig.~\ref{fig2}(h). When the driving pulse intensity is not sufficient to induce a spectral phase variation along the radial direction, no spatial redirection occurs for both the low- and high-density cases as shown in Appendix~\ref{Appen1}. We can thus identify the XUV light-matter interaction at high intensity as a key ingredient to observe a non-trivial off-axis emission pattern.

To demonstrate the robustness of the effect, we proceed with a full calculation to go beyond the few-level description of the system and the uncoupled radial-spectral evolution of the XUV beam. The calculation is performed by solving the three-dimensional (3D) coupled TDSE in the single active electron approximation and the Maxwell wave equation (MWE) in helium gas \cite{PhysRevA.83.013419}. The same XUV laser and medium parameters are employed as before except for an XUV central frequency of 21.1 eV, which is the state energy of the $1s2p$ state in the present TDSE calculations yielded by the used pseudopotential. Moreover, a confocal parameter of 2.5~cm is used, which corresponds to a Gaussian beam waist of 15 $\si{\micro m}$. The calculated spatio-spectral profiles in the far field for the two atomic densities are shown in Figs.~\ref{fig2}(e) and \ref{fig2}(i), respectively. A 0.1-eV shift of the energy axis is implemented to accommodate the energy offset from the preceding value. The result in the low-density case shown in Fig.~\ref{fig2}(e) agrees well with the simple model simulation in Fig.~\ref{fig2}(d), in terms of the single off-axis emission peak at the resonant frequency. However, the spectral profile becomes asymmetric in Fig.~\ref{fig2}(i) as compared to its counterpart from the simple model simulation in Fig.~\ref{fig2}(h). We confirm that this difference between them predominantly originates from the wavefront curvature of the focused beam: A larger XUV confocal parameter in the 3D TDSE-MWE calculation leads to less asymmetric profiles, which agree more with the simple model prediction in which the focusing and defocusing of the XUV beam within the medium are neglected. It represents a breakdown of the simplest uncoupled approximation for the radial phase variation and indicates the spatio-spectral coupling \cite{Akturk_2010} introduced during the realistic interaction scenario, which merits detailed future exploration. Nevertheless, both Figures~\ref{fig2}(h) and \ref{fig2}(i) show clear double-peak structure in the off-axis spatio-spectral profile, which serves as a hallmark for later experimental demonstration of the interplay between intense XUV light-matter interaction and resonant pulse propagation in an optically thick medium.

\begin{figure}
\includegraphics[width=0.5\textwidth]{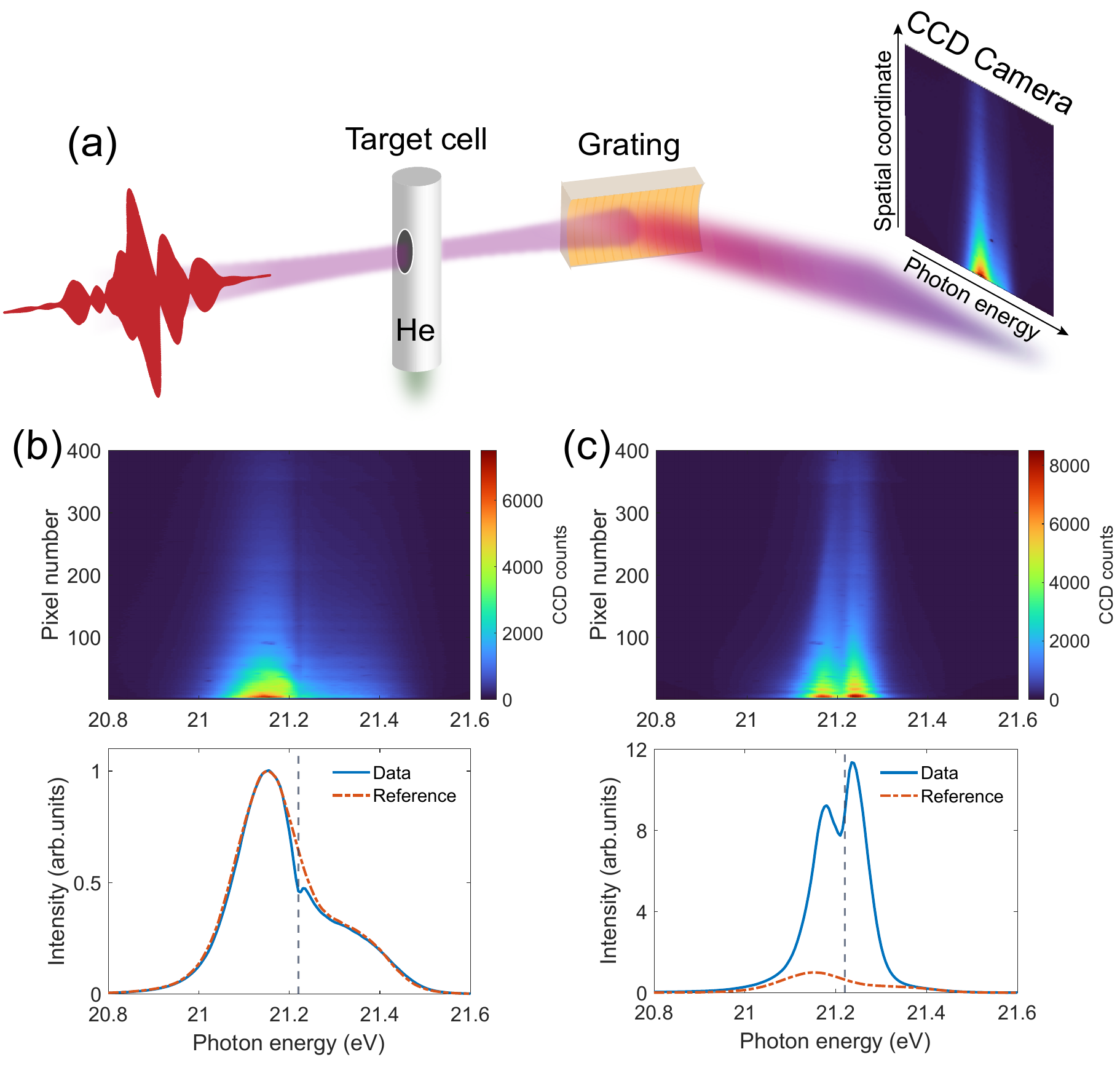}
\caption{\label{fig3} \textbf{Experimental signature of the intense-XUV-induced spatial redirection and spectral reshaping.} (a) Sketch of the experimental setup. The main on-axis part of the dispersed XUV FEL beam was underneath the CCD camera, and only the far off-axis part was experimentally recorded. Recorded 2D CCD image (upper panel) and the spectrum integrated over the whole detected spatial range (lower panel, blue solid line) for the (b) attenuated and (c) unattenuated FEL beam. The results presented in (b) and (c) are averaged over 1400 and 3500 SASE FEL shots, respectively. The vertical dashed line marks the energy position of the resonance. The orange dash-dotted line shows the normalized reference spectrum (shifted in energy) recorded with a central FEL frequency of $\sim$22.1 eV, at which the helium gas is resonance-free and essentially transparent. The depicted spectra are slightly rescaled in amplitude to match the wings of the reference spectrum.}
\end{figure}

Now, we present an experimental signature of the described mechanism by observing the intense-FEL-induced double-peak structure in the transmitted off-axis XUV spectrum. The experiment was performed with the SASE (self-amplified spontaneous emission) FEL pulses at FLASH, DESY \cite{ackermann2007operation}. As illustrated in Fig.~\ref{fig3}(a), the FEL beam was focused into a helium-filled gas cell with an interaction length of 2 mm. The transmitted beam was dispersed and detected in the far field by a two-dimensional (2D) charge-coupled device (CCD) camera, which records the 2D XUV spatio-spectral profile with one spatial axis and one spectral axis. We note that in the experimental configuration, the main on-axis part of the beam was underneath the CCD camera, and only the far off-axis FEL beam profile was recorded. The FEL was operated in single-bunch mode at 10 Hz repetition rate with a pulse energy of $\sim$15~$\si{\micro J}$. The bandwidth of the averaged spectrum was measured to be $\sim$0.2 eV. Considering the beamline transmission of $\sim$30$\%$, the average SASE pulse duration of $\sim$70\textendash90 fs, and the focal beam size of about 10~$\si{\micro m}$, the average on-target peak pulse intensity was in the $10^{13}$~$\si{\rm{W/cm^2}}$ range, which is beyond the perturbative regime \cite{app3010189}. An aluminum filter of 400~nm thickness can be inserted into the beam path to attenuate the incoming beam. Along with the inevitable formation of aluminum oxide layer on the filter surfaces, we estimate an order of magnitude of attenuation in the considered photon energy range of around 21 eV, which is close to the perturbative regime \cite{app3010189}. 

Figure~\ref{fig3}(b) shows the measured spatio-spectral profile on the 2D CCD camera and the spatially integrated spectrum for an attenuated FEL beam with a central photon energy of 21.2 eV. The nominal backing pressure of the gas cell was around 26 mbar. An absorption dip at the resonant position is observed. For the results of the unattenuated FEL beam shown in Fig.~\ref{fig3}(c), an obvious double-peak structure around the resonance position shows up, with an amplitude considerably higher than that of the reference spectrum. This spectral behavior mirrors the theoretical simulations shown before, which confirms the experimental observation of the spatial redirection and spectral reshaping of intense XUV radiation. Hereby, the characteristic off-axis double-peak structure observed in Fig.~\ref{fig3}(c) implies the experimental condition is in the high-XUV-intensity and high-pressure regime, in which both XUV-driven Rabi dynamics and resonant pulse propagation effects are at work.

In summary, we have demonstrated the modification of the spatial and spectral properties of XUV radiation by strongly driving an electronic transition with an intense XUV FEL beam, which translates the radial intensity inhomogeneity into different spectral phase variations in the vicinity of the resonance. The interplay between intense XUV light-matter interaction and resonant pulse propagation in an optically thick medium leads to the redirection and reshaping of the spectral structures. In combination with theory calculations, the characteristic off-axis two emission peaks indicate the presence of XUV-driven Rabi dynamics and macroscopic pulse propagation effects. The described phenomenon is universal and can be exploited as spatio-spectral phase modulators working in the intense short-wavelength regime, where achieving pulse shaping is challenging but promising for element-specific launching and steering electron wavepackets.

We acknowledge DESY (Hamburg, Germany), a member of the Helmholtz Association HGF, for the provision of experimental facilities. Technical contributions were carried out at MPIK and FLASH and we would like to thank both MPIK and FLASH scientific and technical support teams. Beamtime was allocated for proposal F-20220722. Portions of the computation were conducted with high performance computing resources provided by LSU. The work at LSU was supported by the U.S. Department of Energy, Office of Science, Basic Energy Sciences under Contract No. DESC0010431.

\appendix

\section{Simulation results in the low-intensity case}
\label{Appen1}
\begin{figure*}
\includegraphics[width=0.8\textwidth]{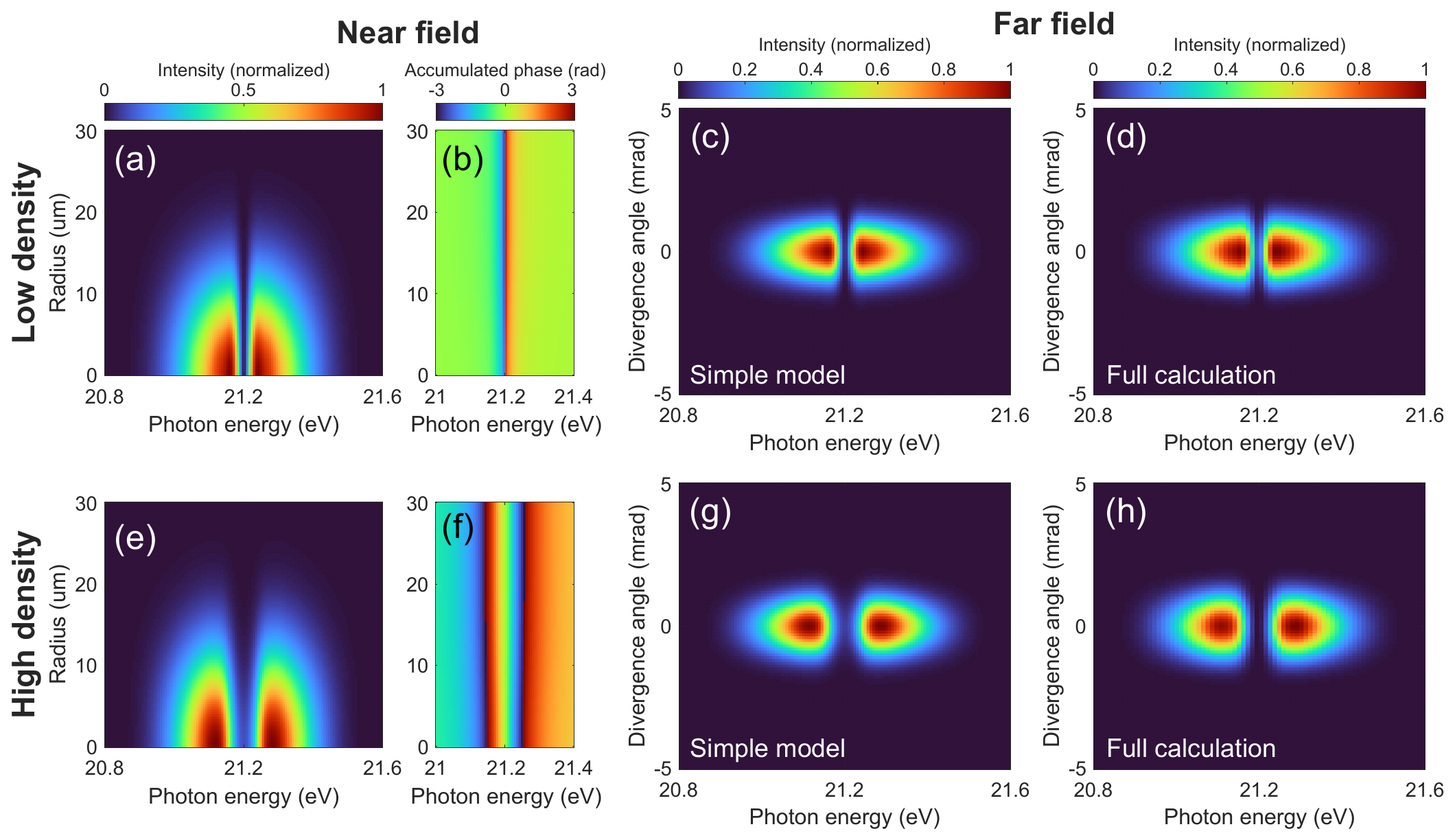}
\caption{\label{figSM} \textbf{Simulation results in the low-intensity case.} (a, e) Near-field radial-spectral profile, (b, f) accumulated phase of the XUV beam at the exit of the medium, and the resulting (c, g) far-field spatio-spectral profiles simulated with the same parameters as in Fig.~\ref{fig2} except for a peak laser intensity of 4 $\rm{TW/cm^{2}}$. (d, h) Far-field spatio-spectral profiles obtained from the full TDSE-MWE simulations. The atomic densities in panels (a-d) and (e-h) are $2.4\times10^{16}$ and $1.9\times10^{17}$ $\si{cm^{-3}}$, respectively.}
\end{figure*}

Figure~\ref{figSM} presents the simulated near- and far-field results with the same parameters as in Fig.~\ref{fig2} in the main text except for a rather low peak XUV pulse intensity of 4~$\rm{TW/cm^{2}}$. Here the accumulated spectral phase of the XUV beam is uniform along the radial direction for both the low- [Fig.~\ref{figSM}(b)] and high-density [Fig.~\ref{figSM}(f)] cases. A spectral dip appears at the transition energy in the far-field spectrum [Figs.~\ref{figSM}(c), \ref{figSM}(d), \ref{figSM}(g), and \ref{figSM}(h)], and no spatial redirection of the XUV beam occurs.

\bibliography{FEL-OffAxisEmission}% Produces the bibliography via BibTeX.

\end{document}